# Aftershocks following crash of currency exchange rate: The case of RUB/USD in 2014

VASILYA USMANOVA[1], YURY V. LYSOGORSKIY[1] and SUMIYOSHI ABE[1,2,3]

[1] *Institute of Physics, Kazan Federal University, Kazan 420008, Russia*

[2] *Physics Division, College of Information Science and Engineering, Huaqiao University, Xiamen 361021, China*

[3] *Department of Physical Engineering, Mie University, Mie 514-8507, Japan*

**Abstract.** The dynamical behavior of the currency exchange rate after its large-scale catastrophe is discussed through a case study of the rate of Russian rubles to US dollars after its crash in 2014. It is shown that, similarly to the case of the stock market crash, the relaxation is characterized by a power law, which is in analogy with the Omori-Utsu law for earthquake aftershocks. The waiting-time distribution is found to also obey a power law. Furthermore, the event-event correlation is discussed, and the aging phenomenon and scaling property are observed. Comments are made on (non-)Markovianity of the aftershock process and on a possible relevance of glassy dynamics to the market system after the crash.

It is known that there exist striking similarities between seismicity, stock markets, and the Internet. Two celebrated empirical laws in seismology, i.e., the Gutenberg-Richter law [1,2] for frequency of earthquakes vs. released energies and the Omori-Utsu law [2-4] for the temporal pattern of occurrence of aftershocks following a shallow main shock, are also observed in stock markets [5,6] and in Ping experiments for the Internet [7,8]. These facts may suggest a possibility that such apparently distinct ones may actually belong to a similar class of complex systems exhibiting catastrophes.

Our purpose here is to study, as another complex system exhibiting catastrophes, a currency exchange market based on the real data analysis. In particular, we focus our attention on the dynamical behavior of the exchange rate of Russian rubles to US dollars (abbreviated as RUB/USD) after the large-scale crash occurred at 20:17 on December 15, 2014, which was followed by a swarm of "aftershocks". In the dataset we employ here (which is available at http://www.finam.ru), the value of RUB/USD is recorded every one minute during the periods of exchange (the periods of no exchange, in which nothing is recorded, are removed from the analysis). We discuss how RUB/USD relaxes to its value in a stationary state. We find that the aftershocks well obey the Omori-Utsu law, implying a very slow relaxation. This bears strong similarity to the case of the stock market crash thoroughly investigated in Ref. [6]. However, an essential difference between them should be noted: the study in Ref. [6] treats the collective behavior of the system (the Standard and Poor's 500 index, specifically), whereas the present one considers only RUB/USD. This means that the behavior of RUB in the global market is reflected in RUB/USD. After identifying the Omori-Utsu law for the exchange rate, we further analyze the waiting-time distribution and correlation between aftershocks. We show that the waiting-time distribution is of the power law, and the event-event correlation function exhibits the aging phenomenon with the associated scaling property. In addition, we also examine if the process of aftershocks is (non-)Markovian.

It seems appropriate to start our discussion with summarizing the Omori-Utsu law [2-4]. Suppose an extreme event (e.g., a shallow main shock in seismicity or a market crash in finance) occurred at time $t=0$. Then, this law states that the number of aftershocks, $\Delta N(t)$, occurring in a time interval $[t, t+\Delta t]$ ($t>0$) asymptotically decays as a power law: $\Delta N(t)/\Delta t = A\,(t+c)^{-p}$, where $p\,(>0)$, $A\,(>0)$, and $c\,(\geq 0)$ are constants. In the continuous approximation, the left-hand side is interpreted as the time derivative of $N(t)$, which is the cumulative number of aftershocks occurred until time $t\,(>0)$

$$N(t) = \begin{cases} A\,[(t+c)^{1-p} - c^{1-p}]/(1-p) & (p \neq 1) \\ A\,\ln(t/c+1) & (p=1) \end{cases}. \qquad (1)$$

The power-law nature of the rate, $\Delta N(t)/\Delta t$, implies that relaxation of the system to its stationary state is very slow.

To define an aftershock, first we consider the following quantity:

$$r(t) = \frac{X(t+\Delta t) - X(t)}{X(t)}, \qquad (2)$$

where $X(t)$ is the value of RUB/USD at $t$ and $\Delta t = 1$ min. This quantity reminds one of the variable termed "price return" of a stock [9,10]. Once a period $[t_0, t_0 + T]$ is fixed, the average and variance are calculated to be

$$\mu = \frac{1}{T} \sum_{t=0}^{T} r(t_0 + t), \qquad (3)$$

$$\sigma^2 = \frac{1}{T} \sum_{t=0}^{T} \left[ r(t_0 + t) - \mu \right]^2, \qquad (4)$$

respectively.

Here, the interval of exchange days after the crash is taken to be 100 days after the moment of the crash of RUB/USD mentioned above. The value of $\sigma$ is calculated to be

$$\sigma = 0.00404, \qquad (5)$$

accordingly.

In Fig. 1, we present the plot of $r(t)$ in Eq. (2) with respect to time. The moment, $t = 0$, is adjusted to be at that of the crash. It exhibits a more intermittent behavior after the crash, compared to the stock-market aftershock regime shown in Ref. [6]. In this respect, the following point should be noted: the Omori-Utsu law for earthquake aftershocks never claims that aftershocks are weaker than a main shock. This behavior may come from piled different complex dynamics at different time scales.

Now, we analyze the cumulative number of "aftershocks" following the crash. We define in our analysis a "shock" or an "event" in such a way that $|r(t)|$ exceeds a certain value of threshold, $R_{th}$. The results in the cases, $R_{th} = 2\sigma$ and $R_{th} = 3\sigma$, are presented in Fig. 2. There, the law in Eq. (1) is seen to hold well. These values of threshold are small compared with those examined in Ref. [6], where the Omori-Utsu law holds well even for $7\sigma$, which is exceptionally large and shows how the financial market crash in 1987 is outstanding. We further note that the values of $p$ here are also small: much smaller than, e.g., unity. (In this respect, we wish to point out the following: in the case of earthquake aftershocks, real data analyses actually show that $p$ varies between about 0.5 and 2.5.)

Next, we consider the waiting-time distribution, $P_W(\tau)$, which is the distribution of the time interval, $\tau$, between successive aftershocks. In Fig. 3, we present the plots of such a distribution for $R_{th} = 2\sigma$ and $R_{th} = 3\sigma$. There, it is seen that the waiting-time distribution well obeys the power law:

$$P_W(\tau) \sim \frac{1}{\tau^{1+\mu}}. \tag{6}$$

This power-law nature is also in strong analogy with seismicity [11,12].

Finally, we discuss the correlation function and its properties. Since the Omori-Utsu law implies nonstationarity of the process of aftershocks, it is of interest to study the nature of correlations. The basic variable employed here is the occurrence time of each aftershock, and a process given by the sequence $\{t_0, t_1, t_2, ..., t_{M-1}\}$ is considered. The label of this sequence, $n \, (= 0, 1, 2, ..., M-1)$, is referred to as *event time*. $t_0$ is the occurrence time of the first aftershock. Then, as in Refs. [13,14], we define the event-event correlation function as follows:

$$C(m,n) = \frac{\langle t_m t_n \rangle - \langle t_m \rangle \langle t_n \rangle}{\sqrt{\sigma_m^2 \sigma_n^2}},  \qquad (7)$$

where the symbol $\langle \bullet \rangle$ denotes the event-time average: $\langle t_m \rangle = (1/M) \sum_{k=0}^{M-1} t_{m+k}$, $\langle t_m t_n \rangle = (1/M) \sum_{k=0}^{M-1} t_{m+k} t_{n+k}$, and $\sigma_m^2 = \langle t_m^2 \rangle - \langle t_m \rangle^2$. To examine (non)stationarity of the process, it is convenient to use the following combination of event times: $C(n+n_w, n_w)$, where $n_w$ is termed *waiting event time*. Clearly, this quantity satisfies the condition, $C(n+n_w, n_w) = 1$ at $n=0$. For a stationary process, $C(n+n_w, n_w)$ does not depend on the waiting event time.

We have analyzed the event-event correlation function defined above for the sequence of the exchange-rate aftershocks. The result is given in Fig. 4. It shows that the process is nonstationary and has the specific property: the larger $n_w$ is, the slower $C(n+n_w, n_w)$ decays, exhibiting the aging phenomenon. In addition, we have also confirmed existence of scaling as the data collapse, as shown in Fig. 5, which implies that

$$C(n+n_w, n_w) = \tilde{C}(n/f(n_w)),  \qquad (8)$$

where, $\tilde{C}$ is a scaling function. From Fig. 6, $f(n_w)$ is seen to be well described by

$$f(n_w) = a(n_w)^\gamma + 1 \qquad (9)$$

where $a$ and $\gamma$ are positive constants.

The aging phenomenon and associated scaling of this type are again in strong analogies with those of earthquake aftershocks [13,14]. However, we note that the interval exhibiting the phenomenon is not long (up to $n \sim 60$). The monotonic behavior ceases to occur for larger values of $n$.

In conclusion, we have studied the dynamical behavior of the currency exchange rate, RUB/USD, just after the crash and have discovered that the "aftershocks" following the crash remarkably share the properties characteristic to earthquake aftershocks including the Omori-Utsu law, power-law waiting-time distribution, and aging phenomenon.

Although we have only looked at RUB/USD, actually this value is linked with a number of other currencies in a complex manner. As can be seen in Fig. 1, the crash has suddenly occurred, and the complex "landscape" of the rates between RUB and other currencies must have been quenched during the crash, analogously to supercooling. Then, the aftershocks have been following. Since the temporal occurrence rate of the aftershocks obey the Omori-Utsu law, which is a power law, relaxation to a stationary state (or, the equilibrium state, if any) is very slow. This observation combined with the existence of the aging phenomenon may suggest that the aftershocks following the large-scale crash of the currency exchange rate is governed by glassy dynamics [15,16], as in the case of earthquake aftershocks [13,14]. However, the "glassy" aspect in the present case is vulnerable, since the regime of aging is not so long. Taking into account all these dynamical features, it is also natural to expect that the sequence of the aftershocks would be a non-Markovian process. One possible way of examining this point is to consider the scaling relation, $p + \mu = 1$, that has to be satisfied by any (singular) Markovian process [17,18]. The condition under which this scaling relation holds is that both $p$ and $\mu$ are in the range, $(0,1)$. It is known [12,14] that this relation is violated for earthquake aftershocks. It seems from Figs. 2 and 3 that the scaling relation is in fact violated also for currency-exchange-rate aftershocks. It is, however, our opinion that further studies are needed to make this point conclusive.

* * *

The authors would like to thank Professor Dmitrii A. Tayurskii for his encouragement and the support from the Program of Competitive Growth of Kazan Federal University by the Ministry of Education and Science of the Russian Federation. SA also thanks for the supports by grants from National Natural Science Foundation of China (No. 11775084) and Grant-in-Aid for Scientific Research from the Japan Society for the Promotion of Science (No. 26400391 and No. 16K05484).

# Figure Captions

Fig. 1  Plot of $r(t)$ in Eq. (2) with respect to time measured every one minute. $t = 0$ is adjusted to be at the moment of the crash.

Fig. 2  Plots of the cumulative number of aftershocks defined in terms of the values of threshold (i) $R_{th} = 2\sigma$ and (ii) $R_{th} = 3\sigma$ with respect to time for 100 currency exchange days after the crash. The solid curves show the model in Eq. (1) with (i) $p = 0.4642$, $A = 6.2409$, $c = 0$ and (ii) $p = 0.4238$, $A = 2.1481$, $c = 0$.

Fig. 3  Log-log plot of the unnormalized waiting time distribution, $P_W(\tau)$ with respect to $\tau$ [min] for the values of threshold (i) $R_{th} = 2\sigma$ and (ii) $R_{th} = 3\sigma$. The histogram is constructed with the bin size, 1 min. The solid lines show the model in Eq. (6) with (i) $\mu = 0.9413$ and (ii) $\mu = 0.9625$.

Fig. 4  Plots of the event-event correlation function, $C(n + n_w, n_w)$, with respect to event time, $n$, for different values of waiting event time, $n_w$: $n_w = $ 0, 10, 20, 30, 40, 50 from the bottom curve to the top. All quantities are dimensionless.

Fig. 5  Data collapse of Fig. 4 through rescaling of event time by $f(n_w)$ to be presented in Fig. 6. All quantities are dimensionless.

Fig. 6  Plot of $f(n_w)$ with respect to $n_w$. The solid line shows the model in Eq. (9) with $a = 0.005$ and $\gamma = 0.984$. All quantities are dimensionless.

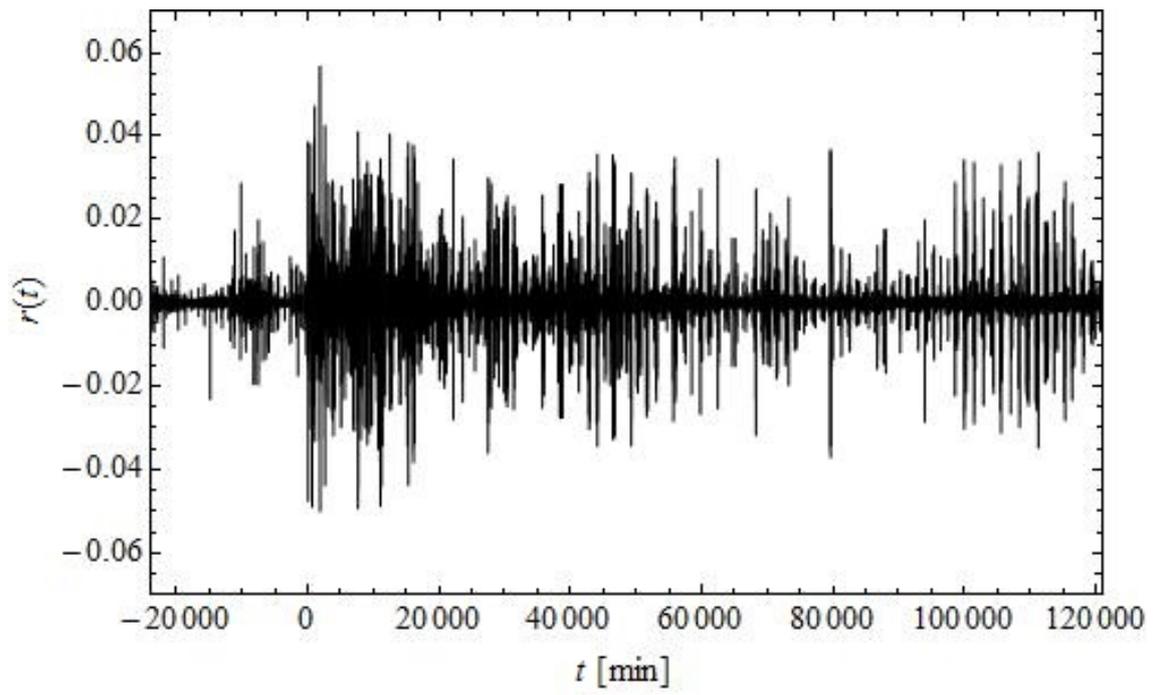

**Fig. 1**

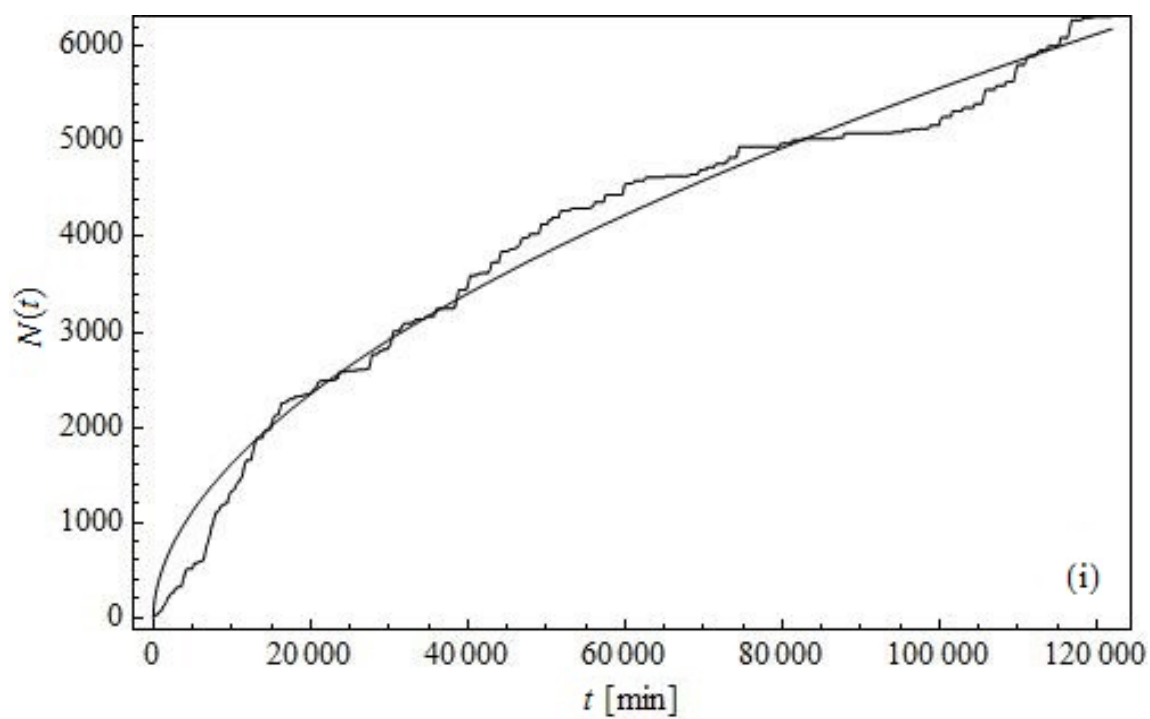

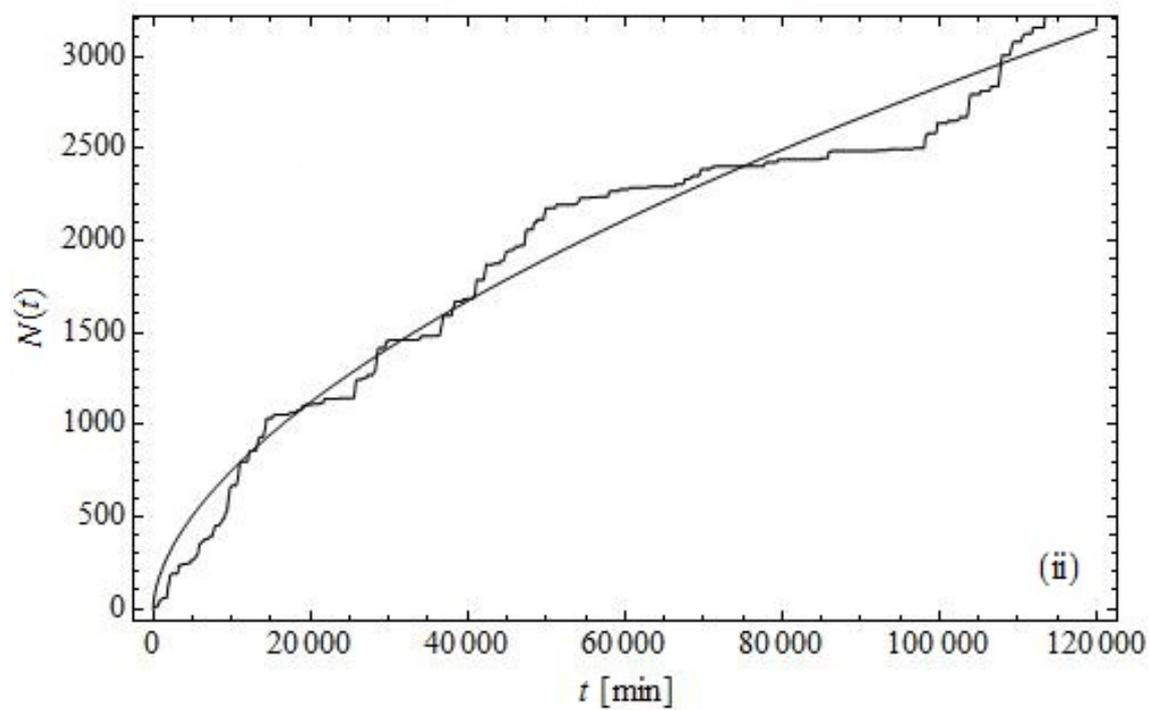

**Fig. 2**

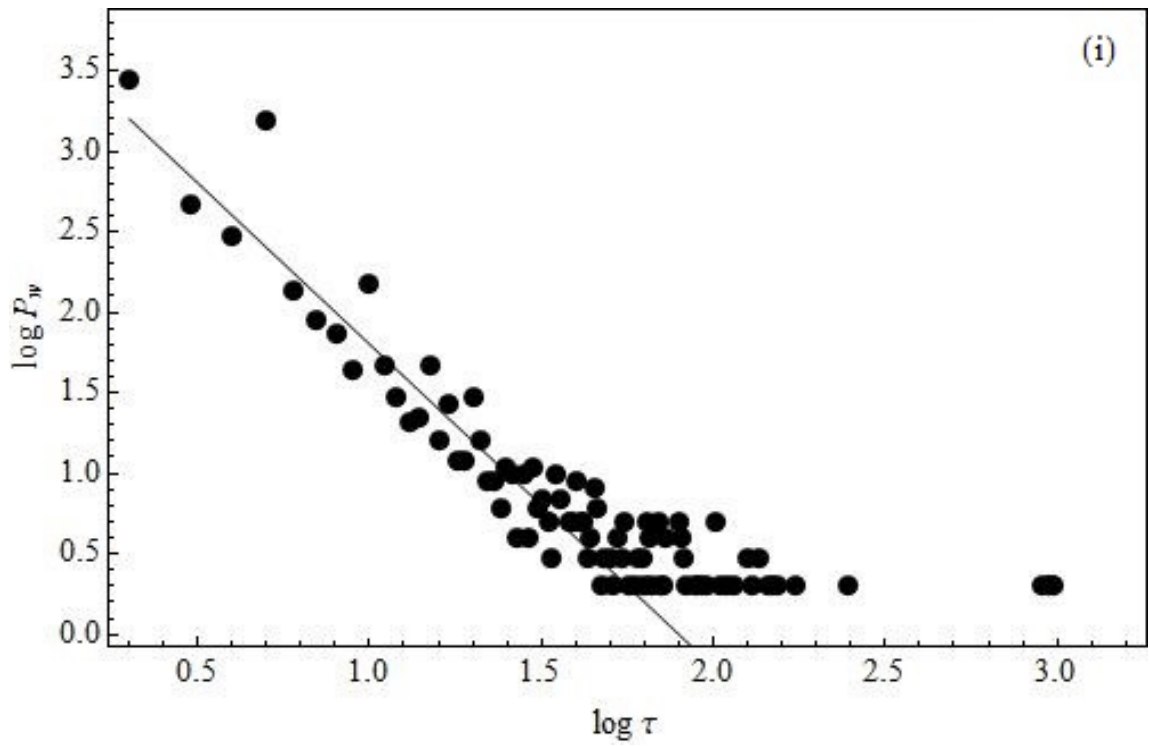

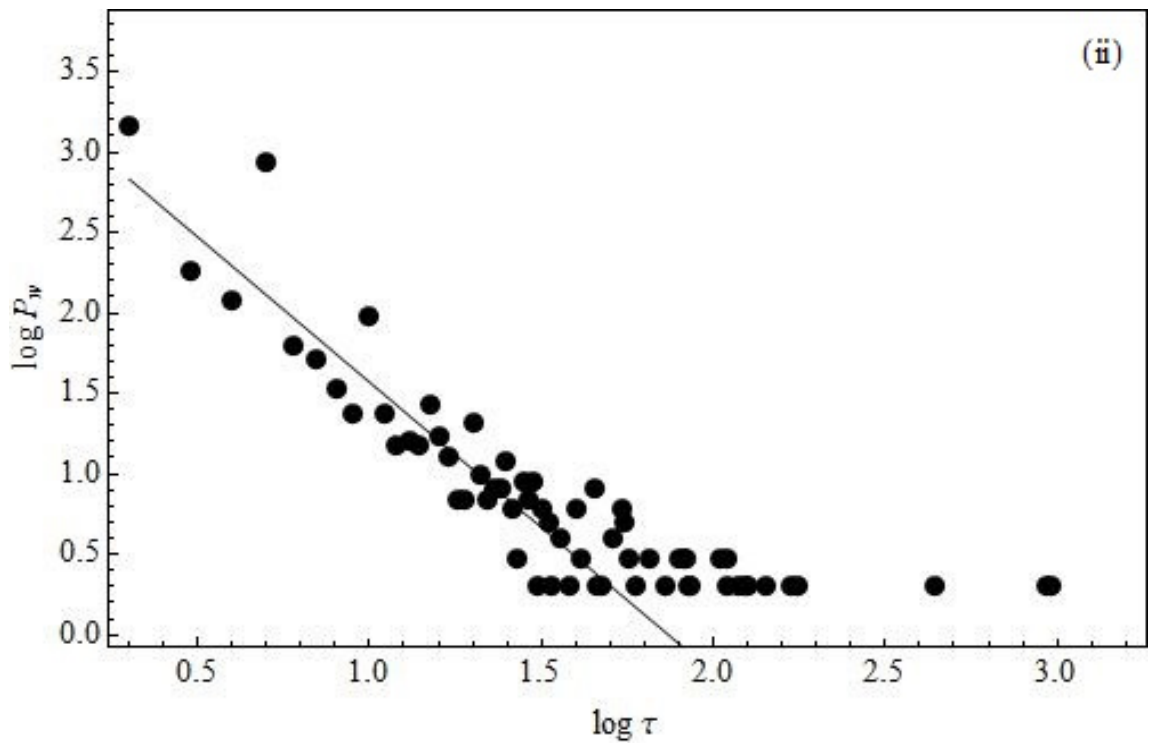

**Fig. 3**

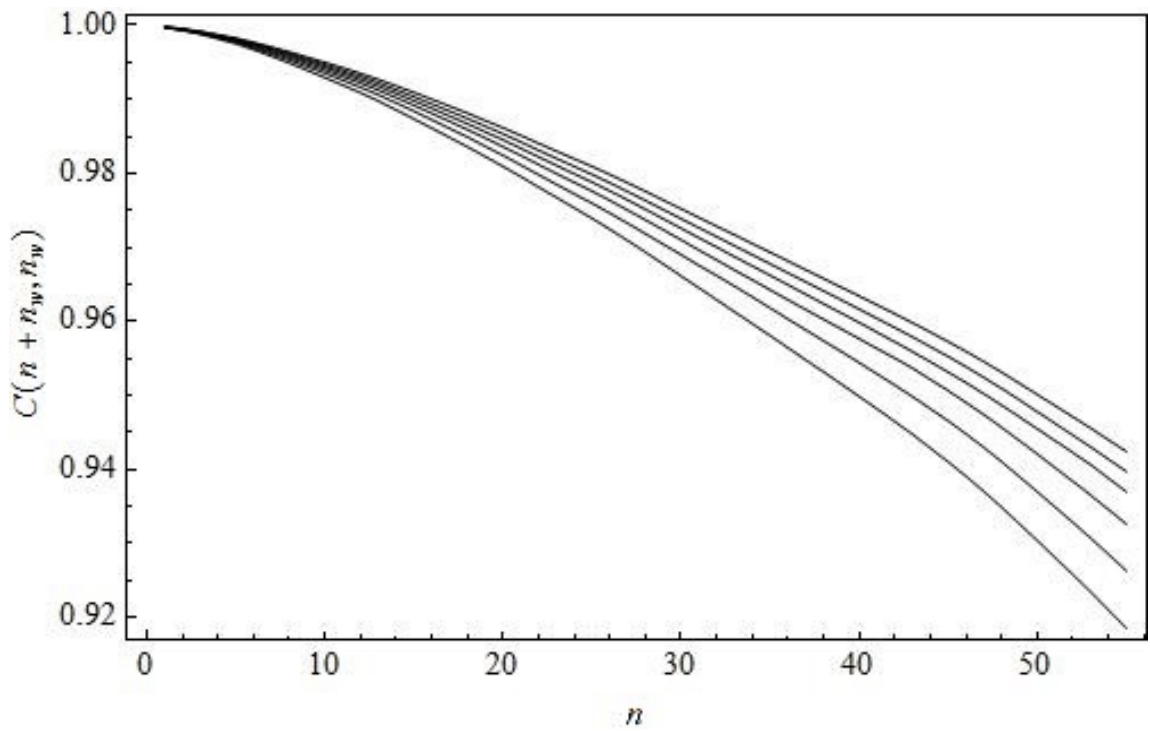

**Fig. 4**

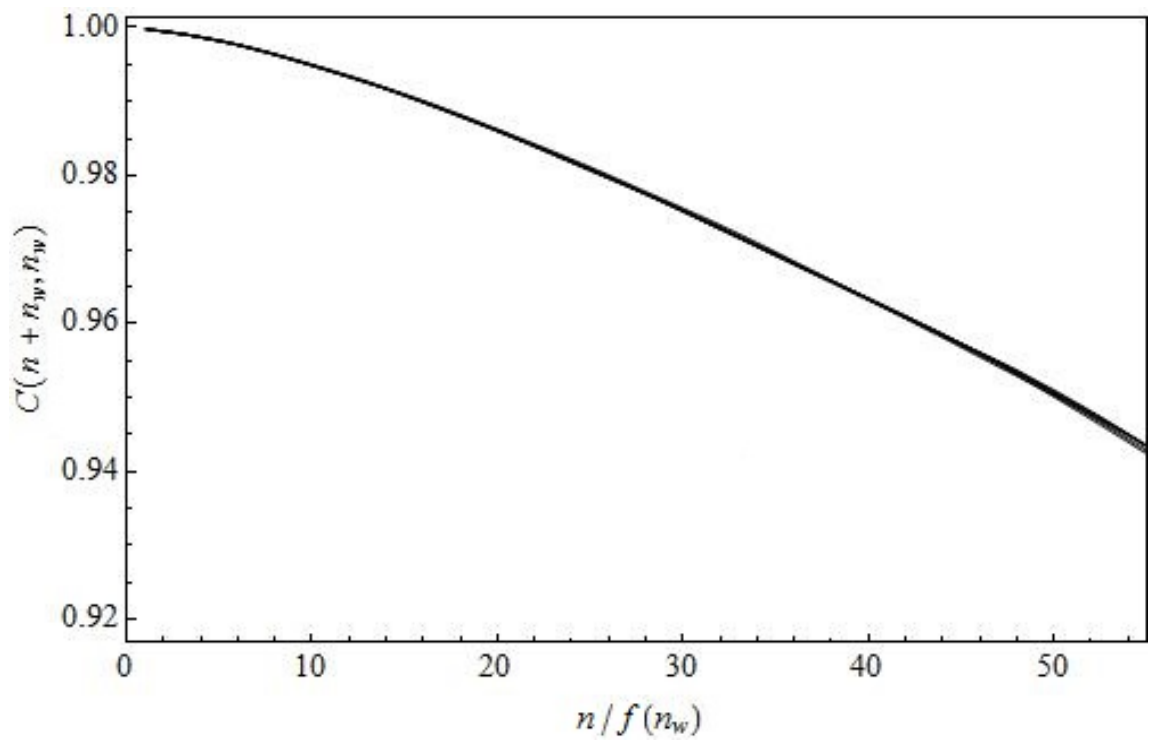

**Fig. 5**

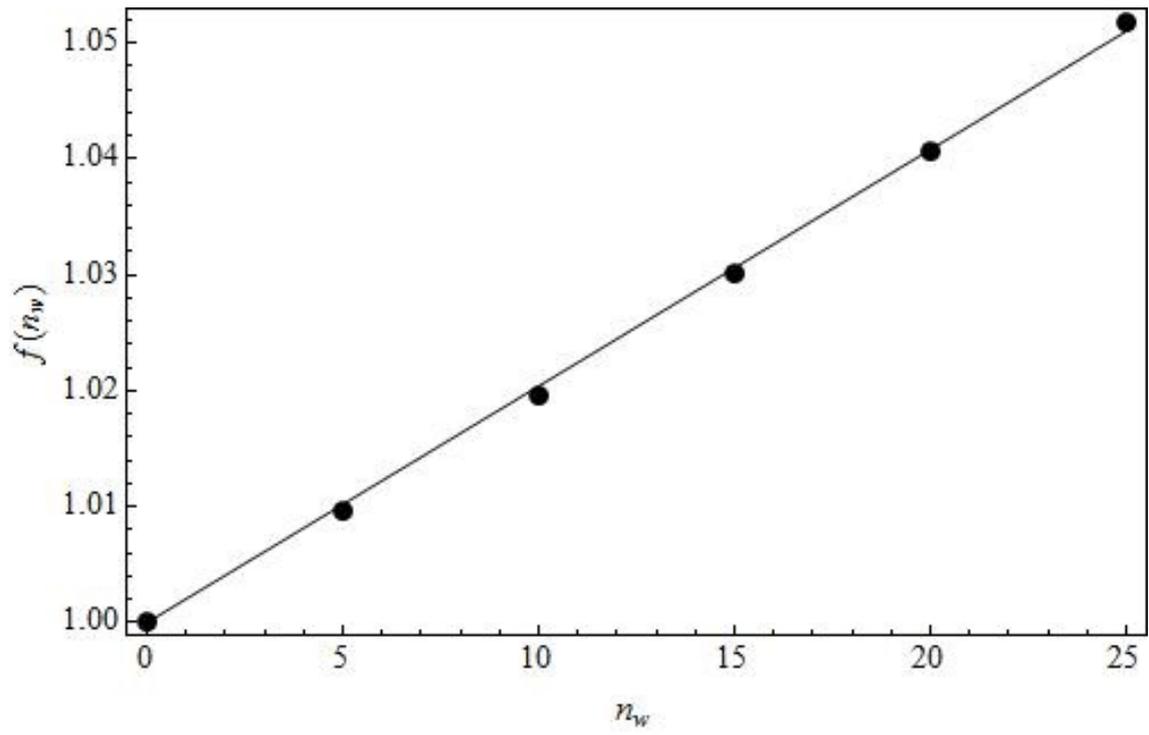

**Fig. 6**